\documentclass[12pt]{article}
\usepackage{amssymb}
\usepackage{graphicx}
\usepackage{psfrag}
\usepackage{amsfonts}
\usepackage{epsfig}
\newcommand{\sef}{\sin^2 \theta_{eff}^{lept}}
\newcommand{\ini}{\begin{equation}}
\newcommand{\fin}{\end{equation}}
\newcommand{\sms}{\hat{s}^2}
\newcommand{\cms}{\hat{c}^2}
\newcommand{\es}{s_{eff}^2}
\newcommand{\ec}{c_{eff}^2}
\newcommand{\drc}{\Delta\hat{r}}
\newcommand{\dre}{\Delta r_{eff}}
\begin{document}

\hyphenation{re-nor-ma-li-za-tion}

\begin{flushright}
NYU-TH/01/02/02\\
hep-ph/0103001
\end{flushright}

\vspace{0.5cm}
\begin{center}
{\Large \bf Scale-Independent Calculation of $\sin^2 \theta_{eff}^{lept}$.}\\
\vspace{0.2cm}
{\large A.~Ferroglia\footnote{e-mail: andrea.ferroglia@physics.nyu.edu}, 
G.~Ossola\footnote{e-mail: giovanni.ossola@physics.nyu.edu}, and A.~Sirlin
\footnote{e-mail: alberto.sirlin@nyu.edu}.}


\vspace{0.5cm}
{\it Department of Physics, New York University,\\
4 Washington Place, New York, NY 10003, USA.}
\end{center}
\bigskip

\begin{center}
\bf Abstract 
\end{center}
{ \small
We present a calculation of the electroweak mixing  parameter $\sin^2 \theta_{eff}^{lept}$
 that incorporates known higher order effects, shares the desirable convergence properties
of the $\overline{MS}$ scheme, and has the important theoretical advantage of being 
strictly independent of the electroweak scale in
finite orders of perturbation theory . We also show how this formulation can 
be extended to the calculation of the $W$ mass $M_W$. The results provide accurate,
scale-independent evaluations of these important parameters, as functions of the Higgs
boson mass $M_H$, and are compared  with previous calculations in order to analyze the 
scheme and scale dependence of the electroweak corrections.}

\newpage
 
The accurate calculation of the electroweak mixing parameter $\sef$ and the $W$-boson
mass $M_W$ rank among the most important objectives of precision studies of the
Standard Model (SM). In fact, these parameters have been measured  very accurately
 and place important constraints on the Higgs boson mass  $M_H$.

Some time ago it was shown that the incorporation of the $\mathcal{O} \left( \alpha^2
M_t^2/ M_W^2 \right)$ contributions \cite{c1} greatly reduces the scheme and 
scale dependence of the radiative corrections,
as well as the upper bound on $M_H$ \cite{c2, c3}. In these papers
the calculations of $\sef$ and $M_W$ were carried out in two implementations 
of the on-shell scheme of renormalization \cite{c4,c5}, denoted as OSI and
OSII, as well as in the $\overline{MS}$ framework \cite{c6,c7,c8,c9,fks}.  Comparison
among the three calculations led then to an analysis of the scheme 
dependence of the electroweak corrections and, by inference, to an estimate
of the theoretical  error arising from the truncation of the perturbative series.
The on-shell and $\overline{MS}$ formulations, which are the most 
frequently employed in electroweak calculations, present a number of
 relative advantages and disadvantages. The on-shell approach  
is very
``physical'', in the sense that it employs renormalized parameters,
such as $\alpha$, $G_\mu$, and $M_W$, that are physical observables
and, therefore, scale independent. The $\overline{MS}$ calculations
follow closely the structure of the unrenormalized theory and, for this reason,
avoid the emergence of large corrections that are frequently induced by 
renormalization. Thus, they have very desirable convergence properties.
On the other hand, they employ parameters, such as $\hat{\alpha} \left( \mu  \right) $
and $\sin^2 \hat{\theta}_W \left(\mu \right)$, which are inherently scale 
dependent. Since $\sef$ and $M_W$ are observable quantities, their 
evaluation, if carried out to all orders in perturbation theory, should lead
 to scale-independent results. Practical calculations, however, involve 
a truncation of the perturbative series and this induces a residual scale
dependence. For instance, among the three schemes discussed in Refs. 
\cite{c2, c3}, only OSII is scale independent.

The aim of the present paper is to present a calculation of $\sef$ that
shares the desirable convergence properties of the  $\overline{MS}$
scheme, but has the important theoretical advantage of being strictly
independent of the electroweak scale
 in
finite orders of perturbation theory. We also show how this formalism can be
extended to the calculation of $M_W$ and applied to the analysis of the
scheme dependence of the electroweak corrections. 

Our starting point is based on two basic relations of the  $\overline{MS}$
renormalization scheme, as applied to electroweak physics \cite{c6, c8, c11}:
\ini \label{eq1}
\sms \cms = \frac{A^2}{M_Z^2 \left( 1- \Delta\hat{r} \right)} \ ,
\fin

\ini \label{eq2}
\es = \left[  1 + \frac{\hat{e}^2}{\sms} \Delta\hat{k} \left( M_Z^2\right)\right]\sms \ ,
\fin
where $A^2 = \pi \alpha / \sqrt{2} G_\mu $, $\sms \equiv  1 - \cms$ and $\es$
are abbreviations for the $\overline{MS}$ parameter $\sin^2 \hat{\theta}_W \left(\mu \right)$
and $\sef$, respectively, and $\Delta\hat{r}$  and $ \left(\hat{e}^2 / \sms\right) 
\Delta\hat{k} \left( M_Z^2\right)$
are the relevant electroweak corrections. As $\es$ is defined in terms of the 
$Z_0 \rightarrow \mathnormal{l} \overline{\mathnormal{l}}$ couplings on resonance,
$\Delta\hat{k}$ is evaluated at $ q^2 = M_Z^2$. Numerically,
 $\left(\hat{e}^2 / \sms\right) \Delta\hat{k}\left( M_Z^2\right)$ is very small,
of $\mathcal{O} \left( 4 \times 10^{-4} \right) $ for $\mu \approx M_Z$.
It is important to note that $ \sms, \hat{e}^2, \drc$ and
 $ \Delta\hat{k}\left( M_Z^2\right)$ are scale-dependent quantities, while
$\alpha, G_\mu, M_Z$, and $\es$, being physical observables, are not.

Since the present knowledge of the irreducible two-loop corrections to $\drc$
and $\left(\hat{e}^2/\sms\right) \Delta\hat{k}\left( M_Z^2\right)$ is restricted to
contributions enhanced by powers $\left( M_t^2 /M_Z^2 \right)^n$ ($n =1,2$),
our strategy to obtain scale-independent expressions is to combine Eq.~(\ref{eq1})
with Eq.~(\ref{eq2}) retaining only such contributions (as well as complete one
loop effects), and expressing the results in terms of scale-independent couplings.

Recalling that at the one-loop level $ \Delta\hat{k}\left( M_Z^2\right)$ depends
only logarithmically on $M_t / M_Z$ \cite{c1, c11} and combining  Eq.~(\ref{eq1})
with  Eq.~(\ref{eq2}), we find

\ini \label{eq3}
\es \ec = \frac{A^2}{M_Z^2 \left( 1- \drc \right) \left[ 1- 
\frac{\hat{e}^2}{\sms} \Delta\hat{k} \left( 1-\frac{\es}{\ec}\right)\right]} \ .
\fin
At the one-loop level \cite{c2, c6, c8}

\begin{displaymath}
1- \drc = 1 + \left. \frac{2 \delta e}{e} \right|_{\overline{MS}} + 
 \frac{e^2}{\sms} \Delta \hat{\rho} +\cdots \quad, 
\end{displaymath} 
where $\left. \frac{ \delta e}{e} \right|_{\overline{MS}}$ is the charge renormalization
counterterm in the $\overline{MS}$ scheme, $\Delta \hat{\rho}$ is given in
Eq.~(4) of Ref. \cite{c2}, and the dots stand for one-loop contributions not
involving the factor $M_t^2 /M_Z^2  $. Noting that $\left(1 +
\left. \frac{2 \delta e}{e} \right|_{\overline{MS}} \right) \hat{e}^2 = e^2 $
\cite{c8} and $\hat{e}^2/\sms = G_\mu 8 M_W^2 / \sqrt{2} + \cdots$ \cite{c2, c8},
where the dots indicate higher order terms, we see that Eq.~(\ref{eq3}) can be 
written in the form 

\ini \label{eq4}
\es \ec = \frac{A^2}{M_Z^2 \left( 1 - \dre \right)} \ ,
\fin

\ini \label{eq5}
\dre = \drc + \frac{e^2}{\es} \Delta \hat{k} \left(1- \frac{\es}{\ec} \right)
\left( 1 + x_t \right) \ , 
\fin
where

\ini \label{eq6}
x_t = \frac{3 G_\mu}{8 \sqrt{2}  \pi^2} M_t^2
\fin
is the leading one-loop contribution to $ \left( \hat{e}^2/ \sms\right) 
\Delta \hat{\rho} $ and we have neglected two-loop corrections not 
proportional to $\left( M_t^2/M_Z^2\right)^n $. The one-loop 
approximation to Eqs.~(\ref{eq4}, \ref{eq5}) has been recently applied
to discuss the mass scale of new physics in the Higgs-less scenario
\cite{c12} and the evidence for electroweak bosonic corrections in the SM
\cite{c13}.

At this stage, it is convenient to express $\drc $ in terms of the corrections
$\drc_W$, $\Delta \hat{\rho} $, and $\hat{f}$ discussed in Refs. \cite{c2, c7, c8}.
Using Eqs.~(15b, 15a, 8b, 8a) of Ref. \cite{c8}, we obtain

\ini \label{eq7}
\drc = \drc_W - \left( e^2 / \es \right) \Delta \hat{\rho} -
\left( e^2 / \es \right) x_t \left[ 2 \Delta \hat{\rho} -
\left( \Delta \hat{\rho} \right)_{lead} - \hat{f} + \Delta \hat{k}\right] \ ,
\fin 
where we have again neglected two-loop contributions not proportional to 
$\left( M_t^2/M_Z^2\right)^n $ and $\left( \Delta \hat{\rho} \right)_{lead} 
= \left( 3 / 64 \pi^2 \right) M_t^2/ M_W^2 $. Combining Eq.~(\ref{eq5})
and Eq.~(\ref{eq7}), we have

\begin{eqnarray} \label{eq8}
\dre & = &\drc_W - \frac{e^2}{\es} \left[\Delta \hat{\rho} -
\Delta \hat{k} \left( 1-\frac{\es}{\ec}\right)
\right]  \nonumber \\
& - & \frac{e^2}{\es} \ x_t \left[ 2 \Delta \hat{\rho} -
\left( \Delta \hat{\rho} \right)_{lead} - \hat{f} + \Delta \hat{k} 
\frac{\es}{\ec}\right] 
 \ .
\end{eqnarray}
The corrections $\drc_W$, $\left( e^2 / \es \right)\Delta \hat{\rho}$
and  $\left( e^2 / \es \right)\Delta \hat{k}$ include irreducible two-loop 
contributions proportional to $\left( M_t^2/M_Z^2\right)^n $ \cite{c1, c2}.
In Eq.~(\ref{eq8}) $\sms$ has been substituted everywhere by the 
scale-independent parameter $\es$. However, $\drc_W$, $\Delta \hat{\rho}$,
 $\Delta \hat{k}$,  and $\hat{f}$ depend also on $c^2 = M_W^2/M_Z^2$,
where $c^2$ is an abbreviation for the on-shell  parameter $\cos^2 \theta_W$
\cite{c4, c5}.
In order to obtain an expression that depends only on $\ec$,  we proceed as
follows: i) $M_W^2$ is replaced everywhere by $c^2 M_Z^2$, 
ii) in all the contributions that are explicitly of two-loop order 
we substitute $c^2 \rightarrow \ec$, since the difference is of third order, and
iii) in the one-loop terms we perform a Taylor expansion, exemplified by
\ini \label{eq9}
\Delta \hat{\rho} \left( \ec, c^2 \right) =   
\Delta \hat{\rho} \left( \ec, \ec \right) + \left.\frac{\partial \Delta \hat{\rho}}
{\partial c^2} \right|_{c^2 = \ec} \left( c^2 - \ec \right) +
\cdots \ .
\fin
In the leading contribution to $\Delta \hat{\rho}$, proportional to $M_t^2$,
$ c^2 - \ec$ in Eq.~(\ref{eq9}) is replaced by the complete one-loop
expression (see Eq.~(\ref{eq13}))

\ini
 c^2 - \ec = \left( G_\mu M_Z^2 / \sqrt{2} \right) 8 \ c_{eff}^4 
\left[ \Delta \hat{\rho} + \left(\es/ \ec \right) \Delta \hat{k}\right] \ ,
\fin
while in the terms not proportional to $M_t^2$
(and this includes $\drc_W$ and $\Delta \hat{k}$) we employ only the 
leading part $c^2 - \ec = \ec \ x_t$. In this way, the calculation of $\es$
is completely decoupled  from that of $M_W$, and can be carried out 
iteratively on the basis of Eq.~(\ref{eq4}) and Eq.~(\ref{eq8}).

In order to evaluate $M_W$, it is convenient to consider the relation
\cite{c8}

\ini
c^2 = \hat{\rho}\  \cms = \left( 1 - \sms \right) \left[
 1 - \left( \hat{e}^2 / \sms\right) \Delta \hat{\rho}\right]^{-1} \ .
\fin
Expressing $\sms$ in the first factor in terms of $\es$ via Eq.~(\ref{eq2}), 
and neglecting again two-loop effects not proportional to
$ \left( M_t^2/M_Z^2\right)^n $, we obtain 

\ini
c^2 = \ec \left\{ 1- \frac{\hat{e}^2}{\sms}\left[ \Delta \hat{\rho}+
\frac{\es}{\ec}  \Delta \hat{k} \left( 1 - \frac{\hat{e}^2}{\sms}
\Delta \hat{\rho} \right) \right] \right\}^{-1} \ .
\fin
Next, we insert the relation 

\ini
\hat{e}^2 / \sms = \left( G_\mu /\sqrt{2}\right) 8 M_W^2
\left[ 1 -  \left( \hat{e}^2 / \sms \right) \hat{f} \right] \ ,
\fin
which follows from Eq.(\ref{eq8}) of Ref. \cite{c2}. This leads to 

\ini \label{eq13}
M_W^2/M_Z^2 = \ec \left\{ 1 - \frac{G_\mu}{\sqrt{2}} 8 M_W^2 \left[
\Delta \hat{\rho} + \frac{\es}{\ec} \Delta \hat{k}
\left( 1 -x_t\right) - \hat{f} x_t \right]\right\}^{-1} \ .
\fin
Since the only one-loop contribution proportional to $M_t^2$ is
$\left( \Delta \hat{\rho}\right)_{lead}$, and this is independent of
$\cms$, this parameter can be replaced by $\ec$ everywhere in the 
expression between curly brackets.
Replacing again $M_W^2 = c^2 M_Z^2$, this expression can be regarded
as a function $G \left(\ec, c^2 \right) $.
Performing a Taylor expansion
about $c^2 = \ec $, analogous to Eq.~(\ref{eq9}), we obtain then a function
that depends only on $\ec $. This fact insures the strict 
electroweak-scale independence of the result and permits the numerical evaluation
of $M_W$ on the basis of Eq.~(\ref{eq13}) and the $\ec$ values 
obtained before.

Since $\es$ plays the role of the basic renormalized electroweak
mixing parameter, for brevity we will refer to the present
framework as  the ``effective'' (EFF) scheme of renormalization.
Numerical results for $M_W$ and $\es$ based on this scheme
are presented in Tables \ref{tab1} and \ref{tab2}, as functions
of $M_H$, using  $G_\mu = 1.16637 \times 10^{-5} \ GeV^{-2}$,
$M_Z = 91.1875 \ GeV$, $\alpha_s \left( M_Z \right) = 0.119$, and 
$M_t = 174.3 \ GeV$  for the pole mass of the top
quark. Table \ref{tab1} employs the traditional value 
$\Delta \alpha_h^{\left( 5 \right)} = 0.02804 \pm
0.00065 $ \cite{jeg} for the five-flavor hadronic contribution to
$\alpha \left( M_Z \right)$, while Table \ref{tab2} is based on
$\Delta \alpha_h^{\left( 5 \right)} = 0.02770 \pm
0.00017 $ \cite{new}, one of the recent ``theory driven'' calculations.
In both Tables, QCD contributions are implemented in the $\mu_t =
\overline{M_t} \left( \mu_t \right)$ scheme explained, for instance,
in Ref. \cite{c2} ($\overline{M_t} \left( \mu_t \right)$ is the
$\overline{MS}$ running top quark mass at  scale $\mu$).
As the dependence of $\Delta \hat{\rho} \left( \ec , c^2 \right)$,
$\Delta \hat{r}_W \left( \ec , c^2 \right)$, $
 \Delta \hat{k} \left( \ec , c^2 \right)$, $\hat{f} 
\left( \ec , c^2 \right)$, and $G \left( \ec, c^2 \right)$ on $c^2$ is rather 
involved, we have found convenient to evaluate numerically, 
rather than analytically, the
derivatives with respect to $c^2$ exemplified in Eq.~(\ref{eq9}).

Calculations in the $\overline{MS}$-scheme traditionally employ $\mu =
M_Z$ when evaluating electroweak corrections at or near the 
$Z^0$-resonance region \cite{c2}. Detailed comparisons to six significant 
figures show that the differences ($\overline{MS}$ at $\mu = M_Z$
minus EFF calculations) are $\delta \es =
\left( 1.6, 1.4, 0.9, 0.6, 0.5\right) \times 10^{-5}$ and
$\delta M_W =
\left( 0.2, 0.3, 0.4, 0.2, -0.1\right) MeV$ for
$M_H = \left( 65, 100, 300, 600, 1000 \right) \ GeV$, respectively,
when $\Delta \alpha_h^{\left( 5 \right)} = 0.02804$ is employed, 
with essentially the same values 
for $\Delta \alpha_h^{\left( 5 \right)} = 0.02770$.
We have also compared the results of the $\overline{MS}$ and EFF
frameworks when the $M_t$ implementation of  the QCD corrections is used 
\cite{c2}. In this case the differences are $\delta \es =
\left( 1.2, 1.0, 0.6, 0.5, 0.4\right) \times 10^{-5}$ and 
$\delta M_W =
\left( 0.1, 0.1, 0.0, -0.3, \right.$ $\left.-0.7\right) MeV$ for 
$\Delta \alpha_h^{\left( 5 \right)} = 0.02804$, and essentially the same for
$\Delta \alpha_h^{\left( 5 \right)} = 0.02770$. Thus, the differences between 
the $\overline{MS}$ ($\mu =M_Z$) and EFF calculations are very small, 
with maximal variations of $\delta \es = 1.6 \times 10^{-5}$ (at
$M_H = 65 \ GeV$ with $\mu_t$-QCD corrections), and $\delta M_W = -0.7 \ MeV$
(at $M_H = 1 \ TeV$ with $M_t$-QCD evaluations).

\begin{figure}[p]
\centering
\psfrag{m}{ {\small $\mu\ \left[ GeV \right]$}}
\psfrag{s}{{\footnotesize $\sef$ }}
\resizebox{12cm}{7cm}{\includegraphics{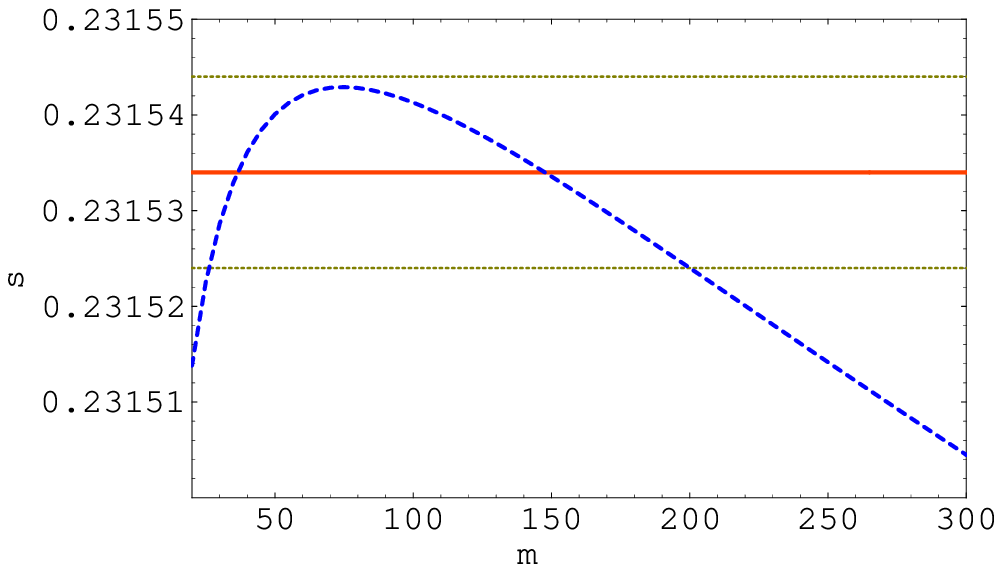}}
\caption{Scale dependence of $\es$ in the $\overline{MS}$ (dashed line)
and EFF (solid line) schemes for $M_H = 100 \ GeV$ 
and the input parameters listed in Table \ref{tab1}.
The light dotted lines define a range of $\pm 1 \times 10^{-5}$ around the EFF result.}
\label{fig1}
\end{figure}

\begin{figure}
\centering
\psfrag{m}{ {\small $\mu\ \left[ GeV \right]$}}
\psfrag{w}{{\small $M_W$} {\small $\left[ GeV \right]$}}
\resizebox{12cm}{7cm}{\includegraphics{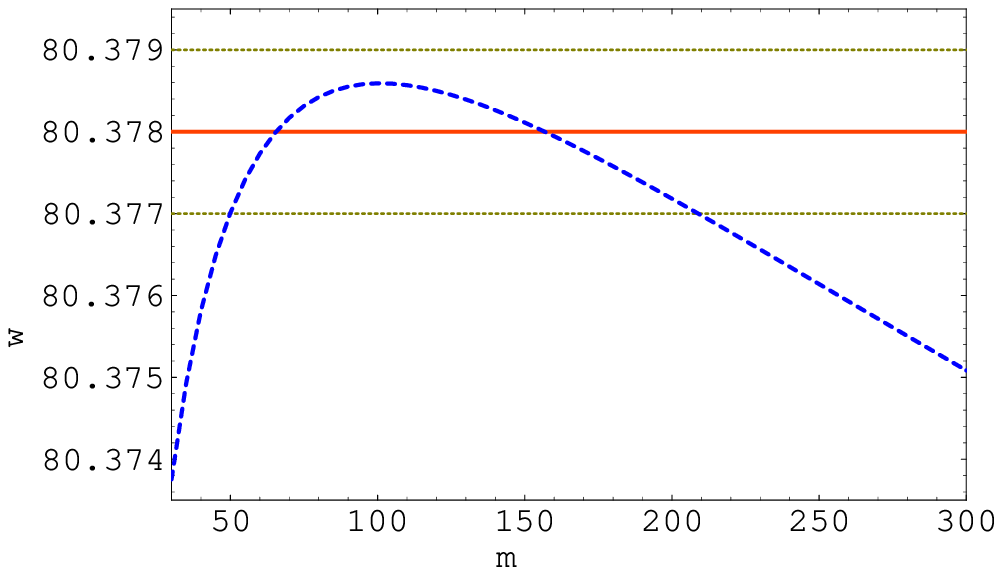}}
\caption{Scale dependence of $M_W$ in the $\overline{MS}$ (dashed line)
and EFF (solid line) schemes for $M_H = 100 \ GeV$ 
and the input parameters listed in Table \ref{tab1}. 
The light dotted lines define a range of $\pm 1 MeV$ around the EFF result.}
\label{fig2}
\end{figure}

Figs. \ref{fig1} and \ref{fig2} compare the $\overline{MS}$ and the EFF 
calculations of $\es$ and $M_W$, as functions of the scale $\mu$,
for $M_H = 100 \ GeV$, $\Delta \alpha_h^{\left( 5 \right)} = 0.02804$,
and $\mu_t$-QCD corrections. We see that the difference $\delta \es$
is within $\pm 1 \times 10^{-5}$ in the range $26 \  GeV \le \mu \le 202 \  GeV$,
but becomes negative and sizable for small and large values of $\mu$. A
very similar pattern holds for the difference $\delta M_W$.
Thus the EFF calculations give support to the $\overline{MS}$ results of 
Refs. \cite{c2, c3}, which employ $\mu =  M_Z$, and at the same time remove the
inherent ambiguities associated with the choice of the electroweak scale.
The effect of such ambiguities on the analysis of the scheme-dependence
is discussed later on.
It is also interesting to note that,
although physicists usually choose   $\overline{MS}$ scales on the basis of
 energies characteristic of the physical observables under consideration, 
 there are important cases in which scales obtained
by optimization methods (BLM \cite{blm}, FAC \cite{fac}, PMS \cite{pms}) are very
different. Recent examples include the relation between pole and  $\overline{MS}$
pole masses, where optimization methods \cite{c21} led 
to extraordinarily accurate predictions 
of third-order coefficients \cite{3ordcoef}, and the QED corrections to 
$\mu$-decay \cite{us}.

A scale-independent estimate of renormalization-scheme differences can be achieved
by comparing the present results with the OSII calculations, 
which are based on the on-shell framework and are also scale independent.
For $\Delta \alpha_h^{\left( 5 \right)} = 0.02804$, $\mu_t$-QCD corrections and 
the same inputs, the differences (OSII minus EFF calculations) are
$\delta \es =
\left( 3.0, 3.3, 3.6, 3.2, 2.3\right) \times 10^{-5}$ and     
$\delta M_W =
-\left( 1.4, 1.4, 1.4, 1.4, 1.2\right) MeV$, for 
$M_H = \left( 65, 100,
300, 600, \right.$ $\left. 1000 \right) \ GeV$. If, instead, 
we employ the $M_t$-QCD implementation, we obtain 
$\delta \es =
\left( 4.2, 4.3, 3.9, 3.1, 1.7\right) \times 10^{-5}$
and $\delta M_W =
-\left( 1.9, 1.8, 1.7, 1.5, 1.1\right) MeV$. 
Comparing the results, for given $M_H$, of the four calculations (OSII
and EFF results, with $\mu_t$ or $M_t$-QCD corrections), 
we find maximal variations of $\approx 5 \times 10^{-5}$ in
$\es$ and $ \approx3 MeV$ in $M_W$.
These conclusions
are very similar to those reported in Ref. \cite{c2}.
For $\Delta \alpha_h^{\left( 5 \right)} = 0.02770$, the maximal 
variations 
are nearly identical to those given above. 
As pointed out in \cite{c3}, the QCD uncertainties are expected to be
larger than indicated by the difference between the $\mu_t$
and $M_t$ approaches, reaching $\pm 3 \times 10^{-5}$
in $\es$ and  $\pm 5 \ MeV$ in $M_W$. Thus, for $M_H = 100 \ GeV$
the overall estimated uncertainty is $\approx 6 \times  10^{-5}$
in $\es$ and  $\approx 7 \ MeV$ in $M_W$.
Using the approximate relations $\delta M_H / M_H \approx 1.9 \times
10^3 \ \delta \es$ and $\delta M_H / M_H \approx -1.6 \times 10^{-2}
\ \delta M_W / MeV$, which can be gleaned from Ref. \cite{c3}, we see 
that the theoretical uncertainties induced by such scheme dependences 
amount to $ \delta M_H / M_H \approx \pm 0.11$  
in both the $\es$ and $M_W$
cases.

It is important to note that, since the $\es$ results of OSII
are larger than those of $\overline{MS}$ for arbitrary
electroweak scale, there is no $\mu$ value for which the two 
calculations coincide and, in fact, their difference increases for
small and large $\mu$ values (cf. Fig. \ref{fig1}).
In contrast, in the $M_W$ case one can choose $\mu$
such that the OSII and $\overline{MS}$ calculations
agree exactly. For instance, for $M_H = 100 \ GeV$,
this occurs at $\mu \approx 45 \ GeV$ and 
$\mu \approx 240 \  GeV$.
This shows that the residual scale ambiguity of the
 $\overline{MS}$ calculations complicates the analysis of 
scheme-dependence and that it is, in fact, highly
advantageous to compare scale-independent calculations,
as we have done above. 

 One may also  compare the $M_W$ calculation in 
the EFF scheme with a recent and more complete on-shell
analysis that incorporates all the two-loop contributions to
$\Delta r$ that contain a fermion loop \cite{hollik}. 
For equal inputs, we find $\delta M_W =
\left( 6.1, 5.3, 2.7, 1.6, 0.7\right) MeV$, ($ \delta M_W \equiv$
EFF minus Ref. \cite{hollik}, using $M_t$-QCD corrections),
with the effective calculation leading to slightly larger $M_H$
values for
 given $M_W$. It is not clear, however, that this comparison 
is a good test of scheme dependence, since Ref. \cite{hollik}
includes a class of two-loop effects not contained in the current
EFF calculation of $M_W$.

As a final application, we list the values of $M_H$ and its
upper-bounds obtained with the EFF calculations on the 
basis of the current experimental values of $\es$ and $M_W$,
without taking into account the experimental lower bounds 
obtained by direct searches.
Using the world average values $\es = 0.23146 \pm 0.00017$,
$M_t = (174.3 \pm 5.1) \ GeV$,
$\hat{\alpha}_s \left( M_Z \right) = 0.119 \pm 0.002$
\cite{leppages}, and $\mu_t$-QCD corrections, we find
$M_H =\left( 86^{+78}_{-41} \ GeV\right)$, $M_H^{95} =
248 \ GeV$ for $\Delta \alpha_h^{\left( 5 \right)} = 0.02804 \pm 
0.00065$, and \mbox{$M_H =\left( 109^{+66}_{-41} \ GeV\right)$}, $M_H^{95} =
238 \  GeV$ for $\Delta \alpha_h^{\left( 5 \right)} = 0.02770 \pm 
0.00016$ ($M_H^{95}$ is the $95 \%$ CL upper bound).
Employing $M_W = \left( 80.436 \pm 0.037\right) \ GeV$ \cite{leppages}
we find $M_H =\left( 28^{+60}_{-28} \ GeV\right)$, $M_H^{95} =
153 \ GeV$ for $\Delta \alpha_h^{\left( 5 \right)} = 0.02804 \pm 
0.00065$, and $M_H =\left( 33^{+62}_{-33} \ GeV\right)$, $M_H^{95} =
162 \ GeV$ for $\Delta \alpha_h^{\left( 5 \right)} = 0.02770 \pm 
0.00016$. We recall that the range $M_H \lesssim 113 \ GeV$ is already 
excluded by direct searches at the $95 \%$ CL. 
It is worth noting that among all  the calculational
schemes we have discussed ($\overline{MS} \left( \mu =M_Z \right)$,
OSII, Ref.\cite{hollik}, and EFF), the latter gives the smallest 
(largest) value of $\es$ ($M_W$), for given $M_H$. Since $\es
\left( M_W \right)$ is a monotonically increasing (decreasing)
function of $M_H$, the EFF calculations lead to the 
largest values of $M_H$.

In summary, we have discussed and implemented a novel
framework of renormalization in which $\es$ plays the role
of the renormalized electroweak mixing parameter. This scheme
shares the desirable convergence properties of the  $\overline{MS}$
approach, with the important theoretical advantage that the 
calculations are strictly independent of the electroweak scale
in finite orders of perturbation theory. Thus, it also 
shares the attractive properties of the on-shell scheme.
When applied to the evaluation of the basic parameters $\es$ and $M_W$,
it leads to results that are very close to those in the $\overline{MS}$
scheme, provided that the scale in the latter calculation is chosen 
in the neighborhood of $M_Z$. Thus, it gives strong support to the 
$\overline{MS}$ calculations carried out in the past and, at the same time,
it removes the ambiguity associated with the choice of 
the electroweak scale. As stressed in the paper, the elimination of this
dependence is important, not only in order to obtain unambiguous 
results, but also in the analysis of the scheme-dependence of the
electroweak corrections.

The authors are greatly indebted to G.~Degrassi and P.~Gambino for 
detailed discussions and access to their computational codes. 
This research was supported by NSF grant PHY-$0070787$.

\begin{table}[p] 
\caption{Predicted values of $M_W$ and $\es$ in the 
 EFF renormalization scheme.
QCD corrections based on $\mu_t$-parametrization. $M_t=174.3 \, GeV$,
$\hat{\alpha}_s \left( M_Z^2\right) = 0.119$, 
$\Delta \alpha_{had}^{\left( 5 \right)} =
0.02804$} \label{tab1}
\begin{center}
\begin{tabular}{ |r|| r  | r |} 
\hline 
 & &  \\
$M_H \left[ GeV\right]$ & $M_w \left[ GeV\right]$  &  $\sef $  \\ 
 & & \\
 
\hline \hline
 65  & 80.401 & 0.23132 \\
100  & 80.378 & 0.23153 \\
300  & 80.305 & 0.23211 \\
600  & 80.251 & 0.23250 \\
1000 & 80.212 & 0.23278 \\
\hline
\end{tabular}
\end{center}
\end{table}

\begin{table}[p] 
\caption{As in Table 1, but with $\Delta \alpha_{had}^{\left( 5 \right)}=
0.02770$.} \label{tab2}
\begin{center}
\begin{tabular}{ |r|| r | r |}
\hline 
 & & \\
$M_H \left[ GeV\right]$ &  $M_w \left[ GeV\right]$ &  $\sef $  \\ 
 & &  \\
 
\hline \hline
 65  & 80.407 & 0.23120 \\
100  & 80.384 & 0.23142 \\
300  & 80.310 & 0.23199 \\
600  & 80.257 & 0.23238 \\
1000 & 80.218 & 0.23266 \\
\hline
\end{tabular}
\end{center}
\end{table}



\end{document}